\documentclass[twocolumn,showpacs,preprintnumbers,nofootinbib,amsmath,amssymb,floatfix,aps,prc,10pt]{revtex4-1}
\usepackage{graphicx}
\usepackage{subfigure}
\usepackage{bm}

\newcommand{\n}[1]{\ensuremath{|\mathbf{#1}|}}
\newcommand{\kPrime}{\ensuremath{|\mathbf{k}'|}}
\newcommand{\pmax}{\ensuremath{p_\text{max}}}
\newcommand{\mmin}{\ensuremath{M^\text{min}_{A-1}}}
\newcommand{\mmax}{\ensuremath{M^\text{max}_{A-1}}}
\newcommand{\emin}{\ensuremath{E_\text{min}}}
\newcommand{\emax}{\ensuremath{E_\text{max}}}
\newcommand{\etal}{{\it et al.}}
\graphicspath{{figs/}}

\begin{document}

\title{Effect of the charged-lepton's mass on the quasielastic neutrino cross sections}
\author{Artur M. Ankowski}
\email{ankowski@vt.edu}
\affiliation{Center for Neutrino Physics, Virginia Tech, Blacksburg, Virginia 24061, USA}

\begin{abstract}
Martini {\it et al.} [Phys. Rev. C {\bf 94}, 015501 (2016)] recently observed that when the pro{\-}duced-lepton's mass plays an important role, the charged-current quasielastic cross section for muon neutrinos can be higher than that for electron neutrinos. Here I argue that this effect appears solely in the theoretical descriptions of nuclear effects in which nucleon knockout requires the energy and momentum transfers to lie in a narrow range of the kinematically allowed values.
\end{abstract}


\maketitle


\section{Introduction}
In a recent article~\cite{Martini:2016eec}, Martini \etal{} analyzed the cross section for charged-current (CC) scattering of electron and muon neutrinos off the carbon target, obtained within two theoretical approaches, the continuum random phase approximation (RPA)~\cite{Jachowicz:2002rr} and the local Fermi gas model with the RPA effects accounted for using the model of Refs.~\cite{Martini:2009uj,Martini:2011wp}. One of the findings of Ref.~\cite{Martini:2016eec} is that at the kinematics where the charged-lepton's mass plays an important role---such as low scattering angles and low neutrino energies---the quasielastic (QE) cross section for muon neutrinos is higher than that for electron neutrinos, contrary to what could be naively expected based on the phase-space availability. Here I point out that this finding is particular to the approaches in which bound nucleons can only be removed from a narrow kinematic region.

\section{Relativistic Fermi gas model}
As an illustrative example, consider the process of CC QE neutrino interaction with a nucleus described within the relativistic Fermi gas (RFG) model. For the energy transfer $\omega$, the kinematically allowed range of the momentum transfer $\n q$ is given by the condition~\cite{Ankowski:2010yh}
\begin{equation}\label{eq:FG}
|h-p_F|\leq \n q\leq h+p_F,
\end{equation}
where $h=\sqrt{(\omega-\epsilon+E_F)^2-M^2}$ and $E_F=\sqrt{M^2 + p_F^2}$, with $p_F$ being the Fermi momentum and $\epsilon$ denoting the separation energy. In numerical calculations, I employ the values $p_F=221$ MeV and $\epsilon=25$ MeV~\cite{Whitney:1974hr}. The minimal energy transfer required for the interaction to happen is $\omega_\text{min}= E^\text{min}_{p'} - E_F + \epsilon$, with $E^\text{min}_{p'}=E_F$ when Pauli blocking (PB) is taken into account and $E^\text{min}_{p'}=M$ when this effect is neglected.

To produce a charged lepton of mass $m$ at the scattering angle $\theta$, a neutrino of energy $E_\nu$ has to probe the allowed swath of the $(\omega, \n q)$ values at
\begin{equation}\label{eq:leptonKin}
\n q = \sqrt{E_\nu^2-2E_\nu \kPrime\cos\theta+\kPrime^2},
\end{equation}
where $\kPrime=\sqrt{(E_\nu-\omega)^2-m^2}$.

The difference between the electron and muon masses, $m_e = 0.51$ MeV and $m_\mu = 105.66$ MeV, can affect the phase space available for quasielastic interaction when (i) the neutrino energy or (ii) the charged-lepton's energy are comparable to $m_\mu$, or when (iii) the scattering angle is sufficiently small.

\begin{figure*}
\begin{center}
\includegraphics[width=0.38\textwidth, viewport = 94 284 447 539]{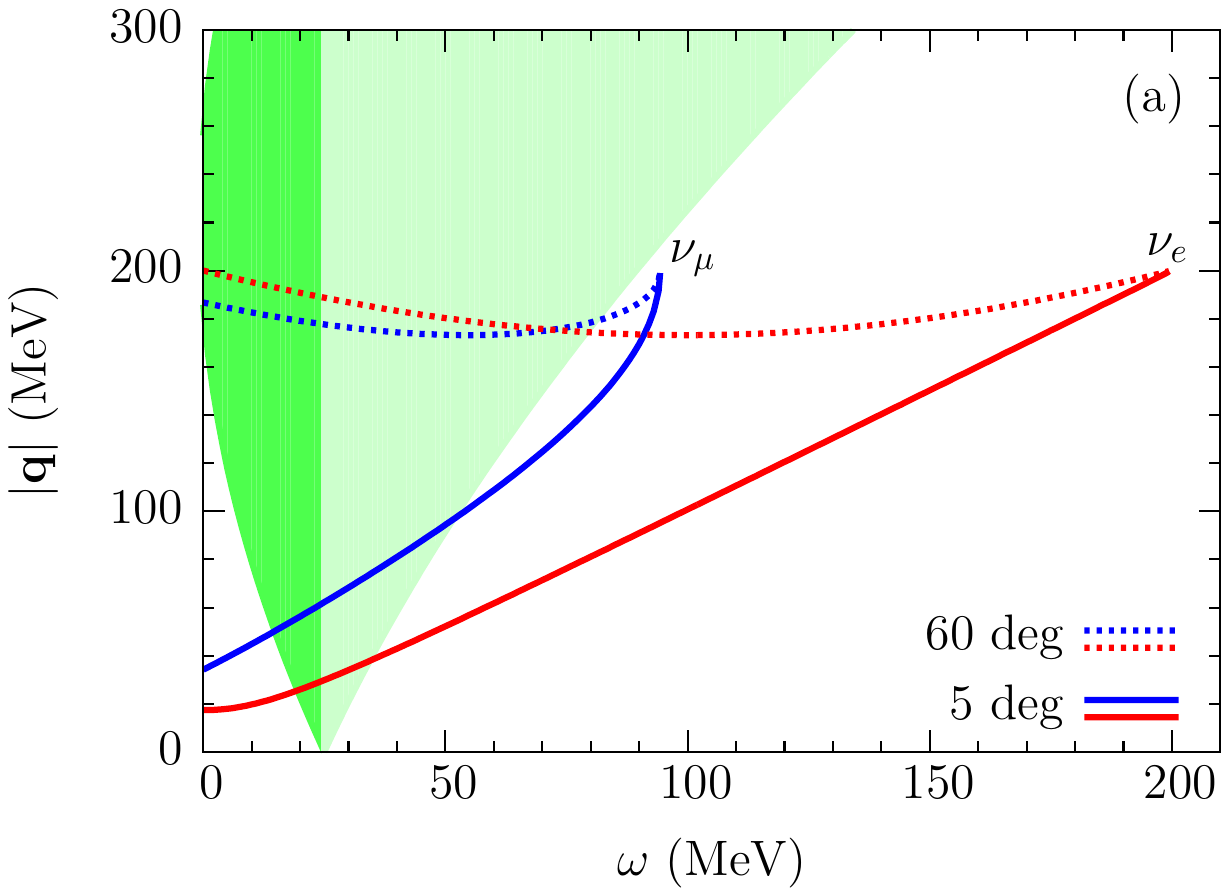}
\hspace{0.8cm}
\includegraphics[width=0.38\textwidth, viewport = 94 284 447 539]{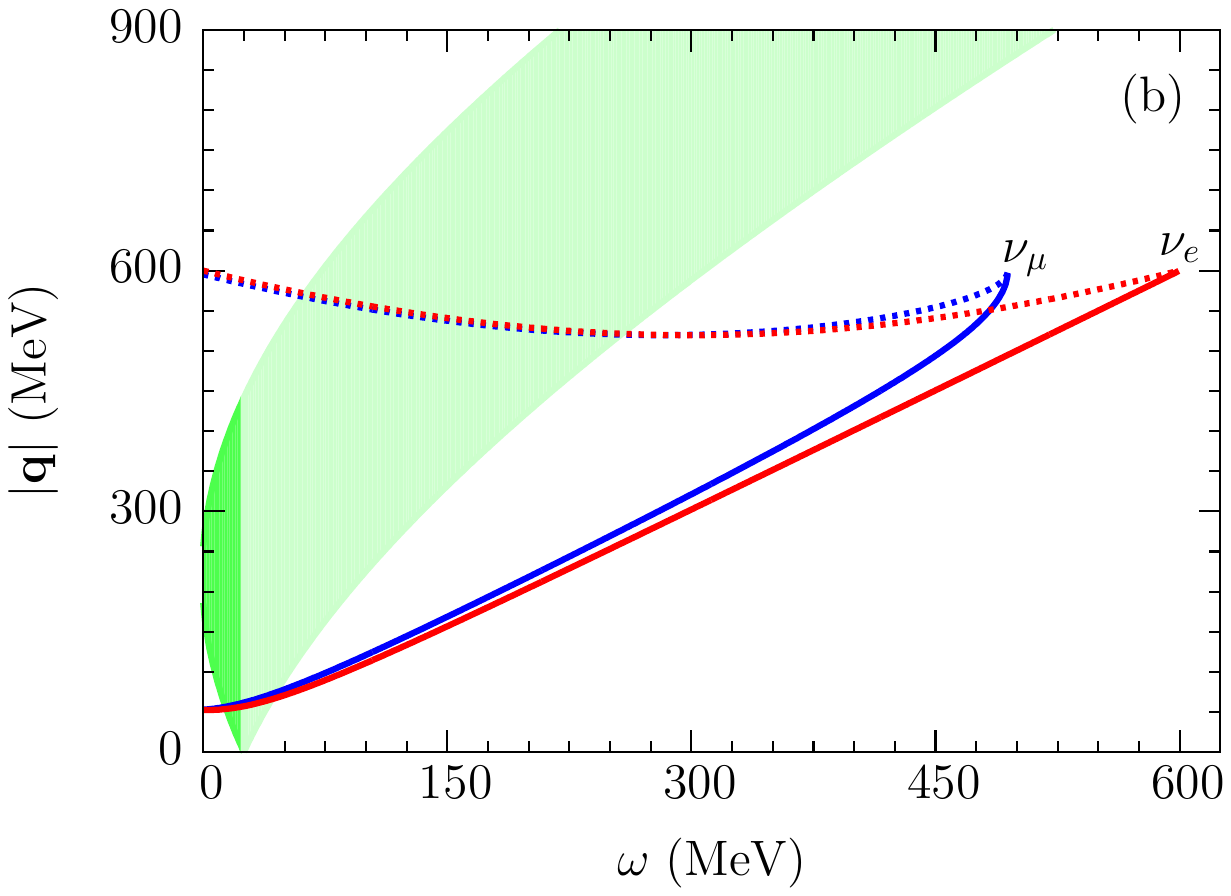}
    \subfigure{\label{fig:FG200}}
    \subfigure{\label{fig:FG600}}
\caption{\label{fig:FG} (color online). Comparison of the values of energy transfer $\omega$ and momentum transfer $\n q$ kinematically allowed in charged-current quasielastic $\nu_e$ and $\nu_\mu$ scattering at $5^\circ$ (solid lines) and $60^\circ$ (dashed lines) for energy (a) 200 MeV and (b) 600 MeV. The light-shaded areas show the $(\omega, \n q)$ regions accessible in the relativistic Fermi gas model with Pauli blocking effect taken into account. The dark-shaded areas represent additional regions available when Pauli blocking is neglected.}
\end{center}
\end{figure*}

These features are shown in Fig.~\ref{eq:FG}, comparing the kinematics accessible for CC QE interactions of neutrinos of energies 200 and 600 MeV and scattering angles $5^\circ$ and $60^\circ$. For low neutrino energies and small scattering angles, the muon-electron mass difference plays an important role and the kinematics probed by electron and muon neutrinos differ sizably, see the solid lines in Fig.~\ref{fig:FG200} corresponding to 200-MeV neutrinos scattering at $5^\circ$. On the other hand, when neutrino energy is significantly higher than the muon mass and the scattering angle is large, electron and muon neutrinos probe the nuclear response at essentially the same values of the energy and momentum transfers; see the dotted lines in Fig.~\ref{fig:FG600} corresponding to 600-MeV neutrinos scattering at $60^\circ$.

\begin{table}[b]
\caption{\label{tab:cs}Ratio of the differential cross sections $\frac{d\sigma(\nu_\mu)}{d\cos\theta}/\frac{d\sigma(\nu_e)}{d\cos\theta}$ for charged-current quasielastic scattering off carbon calculated within different nuclear models.}
\begin{ruledtabular}
\begin{tabular}{l c c c c}
              & \multicolumn{2}{l}{$E_\nu=200$ MeV} & \multicolumn{2}{l}{$E_\nu=600$ MeV}\\
 & $5^\circ$ & $60^\circ$ & $5^\circ$ & $60^\circ$\\
\hline
RFG w/\phantom{o} PB  & 1.57 &  0.62 & 1.03 & 0.97\\
RFG w/o PB            & 0.73 &  0.71 & 0.96 & 0.97\\
Mean-field SF         & 0.72 &  0.53 & 0.96 & 0.97\\
Full SF               & 0.71 &  0.52 & 0.96 & 0.97\\
\end{tabular}
\end{ruledtabular}
\end{table}


\begin{figure*}
\begin{center}
\includegraphics[width=0.38\textwidth, viewport = 94 284 447 539]{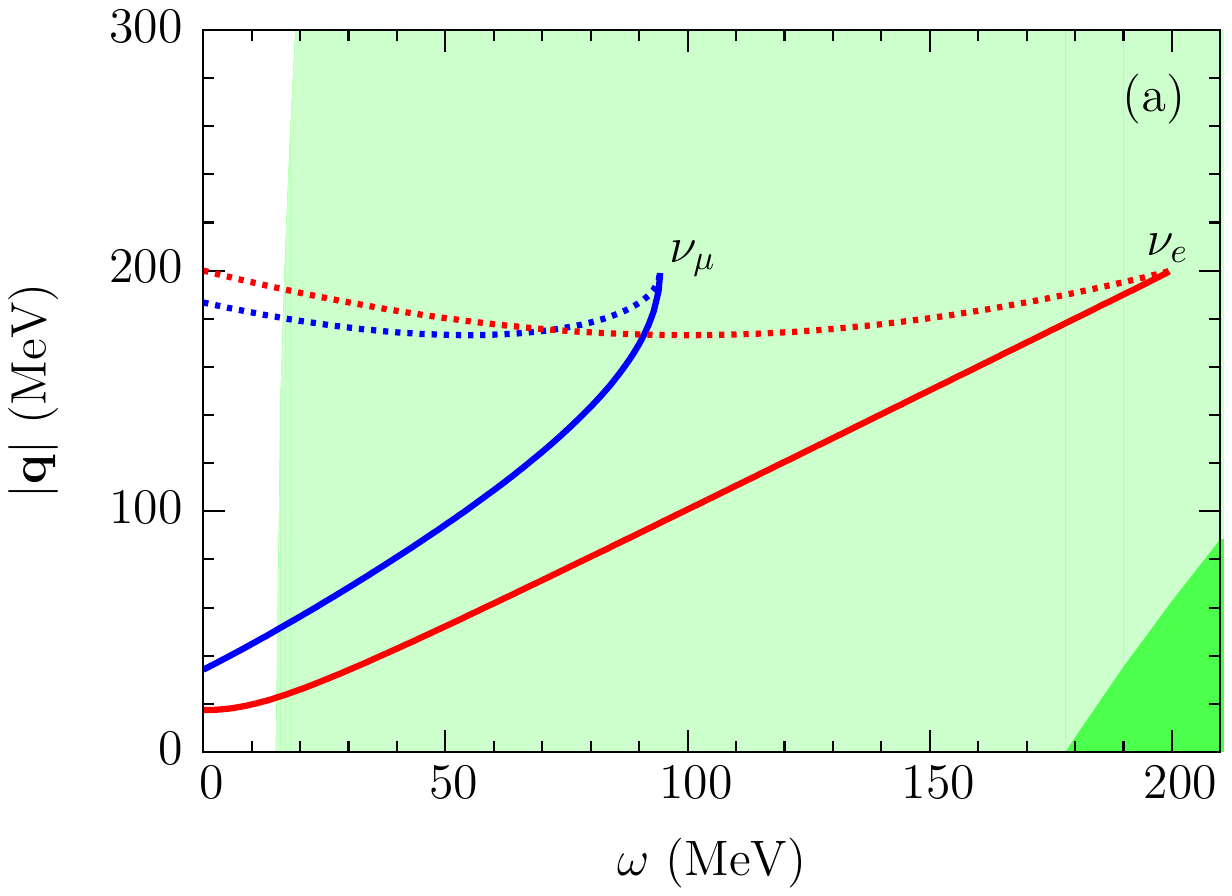}
\hspace{0.8cm}
\includegraphics[width=0.38\textwidth, viewport = 94 284 447 539]{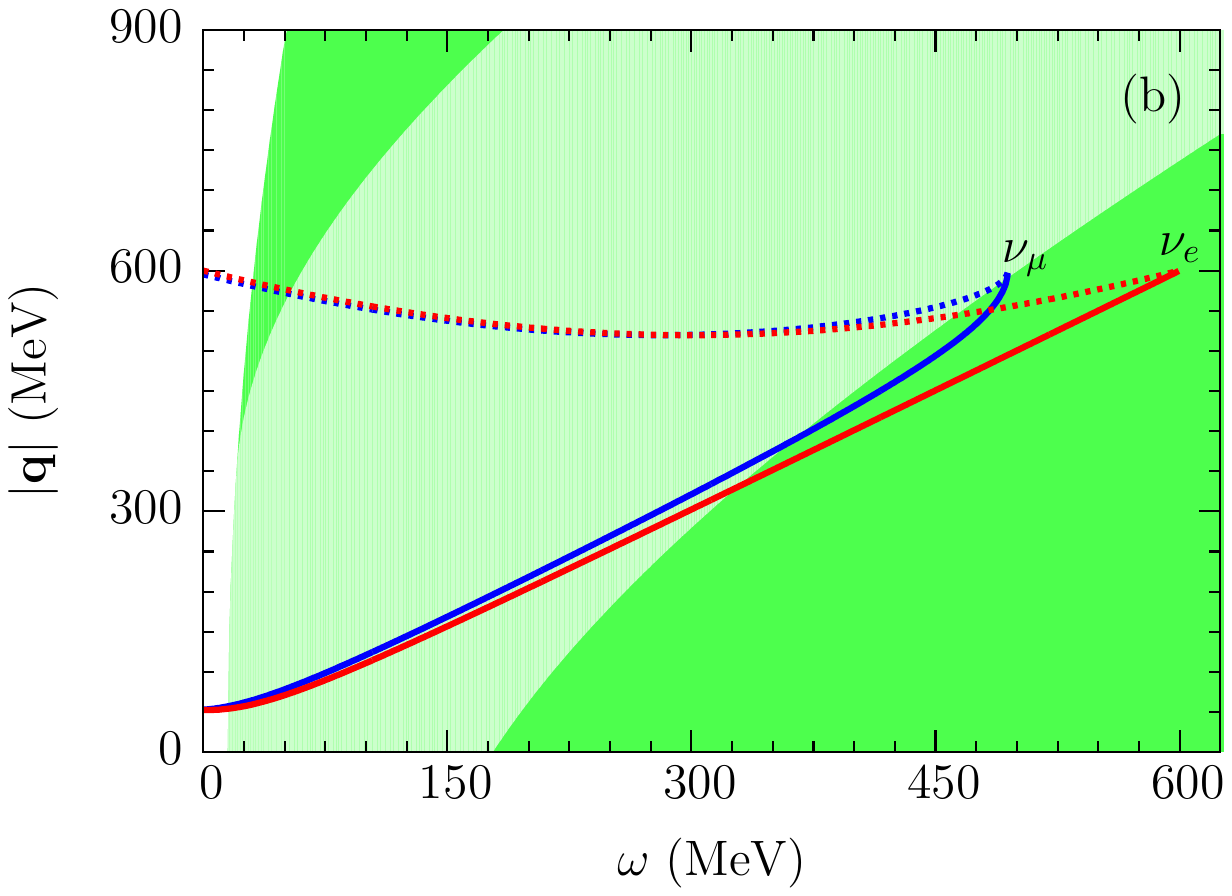}
    \subfigure{\label{fig:SF200}}
    \subfigure{\label{fig:SF600}}
\caption{\label{fig:SF} (color online). Same as Fig.~\ref{fig:FG} but for the spectral function approach. The accessible $(\omega, \n q)$ regions broaden from the light-shaded areas to the dark-shaded ones when the full spectral function is considered instead of its mean-field part only.}
\end{center}
\end{figure*}

At small angles, the RFG model with Pauli blocking yields a CC QE cross section that is higher for muon neutrinos than for electron neutrinos, see Table~\ref{tab:cs}. It is a consequence of the strong constraint that this approach imposes on the allowed values of $\omega$ and $\n q$, clearly visible in Fig.~\ref{fig:FG200}, which affects electron neutrinos much more significantly than muon neutrinos.

However, when the effect of Pauli blocking is neglected, the accessible phase space broadens by gaining an additional area in the $(\omega, \n q)$ space, represented by the dark-shaded regions in Fig.~\ref{fig:FG}. As a result, the CC QE cross section of muon neutrinos becomes lower than that of electron neutrinos, even at low energies and small scattering angles, see Table~\ref{tab:cs}.

\section{Spectral function approach}
The same holds true when the ground-state properties of the carbon nucleus are described using the realistic spectral function (SF)~\cite{Benhar:1989aw,Benhar:1994hw} or even only its mean-field (shell-model) part~\cite{Mougey:1976sc,Dutta:2000sn}, as shown in Table~\ref{tab:cs}. Also in these cases, the broad swath of the allowed $(\omega, \n q)$ values, presented in Fig.~\ref{fig:SF}, ensures that the $\nu_e$ CC QE cross section is not suppressed. At small scattering angles, the interaction process is dominated by low energy transfers, see Supplemental Material~\cite{SupplementalMaterial}, and is largely unaffected by the short-range correlations taken into account in the estimate of the full SF. As a consequence, the cross section ratio $\frac{d\sigma(\nu_\mu)}{d\cos\theta}/\frac{d\sigma(\nu_e)}{d\cos\theta}$ does not change significantly when only the mean-field SF is applied in the calculations.\\[1cm]

\section{General case}
Consider the general case, in which removal of a nucleon of the initial momentum $\n p$, $0\leq \n p \leq \pmax$, leaves the residual $(A-1)$-nucleon system with the excitation energy $E$, $\emin\leq E\leq \emax$, with negligible final-state interactions. The mass of the residual system can vary between $\mmin=M_A-M+\emin$ and $\mmax=M_A-M+\emax$. Then, the kinematically allowed range of the momentum transfer is $|{\bf q}|_\text{min}\leq \n q\leq |{\bf q}|_\text{max}$, with
\begin{widetext}
\begin{equation}
\begin{split}
|{\bf q}|_\text{min}=&
\begin{cases}
0 &\text{for $\emin\leq\omega\leq E_{A-1}^\text{max}+\sqrt{M^2+\pmax^2}-M_A$,}\\
\sqrt{(\omega+M_A-E^\text{max}_{A-1})^2-M^2}-p_\text{max} &\text{for higher $\omega$,}
\end{cases}\\
|{\bf q}|_\text{max}=&
\begin{cases}
\sqrt{(\omega+M_A)^2-(\mmax+M)^2} &\text{for $\emin\leq\omega\leq E_{A-1}^\text{min}\left(1+{M}/{M_{A-1}^\text{min}}\right)-M_A$,}\\
\sqrt{(\omega+M_A-E^\text{min}_{A-1})^2-M^2}+p_\text{max} &\text{for higher $\omega$,}
\end{cases}
\end{split}
\end{equation}
\end{widetext}
where $E^\mathcal{X}_{A-1}=\sqrt{(M^\mathcal{X}_{A-1})^2+p_\text{max}^2}$, $\mathcal{X}=\text{min},\, \text{max}$.\\[-0.8mm]

The broader the allowed ranges of $\n p$ and $E$, the broader the allowed swath of the $(\omega, \n q)$ values.
However, for CC QE scattering of 200-MeV neutrinos at 5 degrees, the ranges $0\leq \n p\leq 240$ MeV and $15\leq E\leq 35$ MeV are sufficient for the maximal allowed energy transfer for $\nu_e$'s to be higher than that for $\nu_\mu$'s. This feature results from lack of one-to-one correspondence between the initial momentum of the struck nucleon and the excitation energy of the residual system, which sizably broadens the allowed kinematics in the region of low energy and momentum transfers, the most significant for scattering at small angles.

\section{Summary}
In the theoretical approaches in which the kinematics of nucleon knockout is strongly constrained---such as the RFG model with Pauli block{\-}ing---the CC QE cross section of muon neutrinos can, in fact, be higher than that of electron neutrinos at small scattering angles, as observed by Martini \etal{}~\cite{Martini:2016eec}. However, this effect does not appear when the allowed swath of the energy and momentum transfers is broader. This is the case, for example, when a theoretical model is able to cover the shell structure determined from the available experimental data~\cite{Mougey:1976sc,Dutta:2000sn} and account for the observed lack of one-to-one correspondence between the momentum of the struck nucleon and the excitation energy of the residual nucleon system.

\acknowledgments
The author thanks Omar Benhar for his comments on the manuscript.
This work has been supported by the National Science Foundation under Grant No. PHY-1352106.

\end{document}